\documentclass[journal,comsoc]{IEEEtran}
\usepackage[T1]{fontenc}
\usepackage[utf8]{inputenc}

\usepackage{mathtools}
\usepackage{amsfonts}
\usepackage{amsmath}
\usepackage{amsthm}
\usepackage{algorithm}
\usepackage[noend]{algpseudocode}

\usepackage{array}
\usepackage{booktabs}
\usepackage{tabularx}
\usepackage{multirow}

\usepackage{xcolor}
\usepackage{acronym}
\usepackage{paralist}
\usepackage{tikz}
\usetikzlibrary{matrix,positioning,shapes.geometric, patterns}
\usepackage{pgfplots}
\usepgfplotslibrary{groupplots}
\pgfplotsset{plot coordinates/math parser=false}
\usepackage[font=footnotesize,labelfont=bf]{caption}
\usepackage{subfig}
\usepackage{graphicx}
\usepackage{interval}
\usepackage[T1]{url}

\newcolumntype{L}[1]{>{\raggedright\let\newline\\\arraybackslash\hspace{0pt}}m{#1}}
\newcolumntype{C}[1]{>{\centering\let\newline\\\arraybackslash\hspace{0pt}}m{#1}}
\newcolumntype{R}[1]{>{\raggedleft\let\newline\\\arraybackslash\hspace{0pt}}m{#1}}

\ifCLASSINFOpdf
  \graphicspath{{figures/}}
  \DeclareGraphicsExtensions{.pdf}
\else
  \graphicspath{{figures/}}
  \DeclareGraphicsExtensions{.eps}
\fi

\usepackage{cite}

\definecolor{bblue}{HTML}{4F81BD}
\definecolor{rred}{HTML}{C0504D}
\definecolor{ggreen}{HTML}{9BBB59}
\definecolor{ppurple}{HTML}{9F4C7C}

\begin{document}

\bstctlcite{IEEE_bibliography_control}

\title{Message-Passing Neural Networks \\ Learn Little's Law}

%
%
%


\author{Krzysztof Rusek and Piotr Cho\l{}da, \IEEEmembership{Member,~IEEE,}
\thanks{AGH University of Science and Technology, Department of Telecommunications, Krakow, Poland. E-mail: krusek@agh.edu.pl.}}

\markboth{IEEE Communications Letters}%
{Rusek and Cho\l{}da: Message-Passing Neural Networks \\ Learn Little's Law}

\IEEEoverridecommandlockouts
\IEEEpubid{\makebox[\columnwidth]{10.1109/LCOMM.2018.2886259~\copyright2018 IEEE \hfill} \hspace{\columnsep}\makebox[\columnwidth]{ }}

\maketitle
\IEEEpubidadjcol


\begin{abstract}
	The paper presents a solution to the problem of universal representation of graphs exemplifying communication network topologies with the help of neural networks. The proposed approach is based on \textit{message-passing neural networks} (MPNN). The approach enables us to represent topologies and operational aspects of networks. The usefulness of the solution is illustrated with a case study of delay prediction in queuing networks. This shows that performance evaluation can be provided without having to apply complex modeling. In consequence, the proposed solution makes it possible to effectively apply methods elaborated in the field of machine learning in communications.
\end{abstract}


\IEEEpeerreviewmaketitle

\section{Introduction, Rationale, and Related Work} \label{sec:intro}

In recent years, we have witnessed an impressive development of machine learning (ML) methods applied in various fields, including network design and management~\cite{klaine17,tutorialmi}. In fact, the importance of these methods is growing, and it is expected that machine intelligence will become the basis for the recently postulated \emph{knowledge plane}~\cite{Mestres:2017:KN:3138808.3138810}. The new plane will be an equal resident of the communication and computer networks (COMNETs), 
with the same rights as in the case of switching, routing, topology planning, security adjustment, etc. This way, network operations will be enriched with automated planning and reaction tools based on ML. Some aspects of this attractive vision are already here, including numerous algorithms which learn and apply the effects of learning, with the most prominent group represented by artificial neural networks (ANNs). They have been used to develop numerous applications, for instance related to image or audio processing, where deep learning enables us to effectively recognize patterns, predict future values, classify behaviors, etc.

Unfortunately, one important element of the network knowledge plane is still missing. When we consider image processing, we can see that there are no problems related to the use of various formats or sizes, blurred pictures and the like. Almost any image can be transformed into a format which can be processed by the assumed ML tool. The case with COMNETs is different. 
We do not have a universal method of representing COMNET structures comparable to those applied in image or sound processing. In this paper, we propose such a general approach by showing how to universally represent COMNETs with ANNs. In consequence, the output ANN representation can be used as a module for various ML applications.

The key issue with COMNET representation is related to its topology, which is typically modeled with graph theory tools. Some approaches to graph representations based on ANNs, not directly focused on COMNETs, were proposed in~\cite{graph_nn}. Moreover, the chemistry field has recently experienced similar difficulties with universal ANN mapping of objects it is interested in. The problem of molecule representation has been successfully overcome using \emph{message-passing neural networks} (MPNN)~\cite{graph_conv,nnmp:17}. Our solution is based on the same ANN architecture. Additionally, we extend the idea by applying batch normalization and using modern SELU (\textit{Scaled Exponential Linear Unit}) activation functions, known for self-normalization and good scaling properties~\cite{Klambauer2017}. It should be noted that the use of MPNNs was proposed recently to support routing in data networks~\cite{Geyer18a}. While we focus on the global performance of the network, the authors of~\cite{Geyer18a} deal with the local (next-hop) behavior.

Section~\ref{sec:idea} outlies the 
solution. Its usefulness is demonstrated by a numerical evaluation based on delay prediction for queuing networks 
in Section~\ref{sec:example}. Section~\ref{sec:conclusions} summarizes the contribution and proposes applications of our approach.

\section{Application of MPNNs}
\label{sec:idea}

First, we elaborate on how message-passing neural architecture can be used for universal representation of graphs. We then show how to adopt this representation in 
COMNETs.

\subsection{Message-Passing Neural Architecture}

Neural message-passing is an ANN architecture originally developed in the form presented here for quantum chemistry~\cite{nnmp:17}. A single message-passing neural network is constructed for any graph. In consequence, the information contained in the graph (including the topology) is compressed as a vector of real numbers. To define the architecture of MPNNs, the following assumptions are made:
\begin{inparaenum}[\bfseries \itshape (a)]
	\item Any structure (a molecule, COMNET, etc.) is represented by a digraph $G = (V,E)$.
	\item Each vertex/node $v\in{V}$ is characterized with an arbitrary set of features (related to topology and also operational aspects), represented by vector $\mathbf{x}_v$.
	\item Likewise, each arc $(v,w)\in{E}$ is characterized with an arbitrary set of features, represented by vector $\mathbf{e}_{vw}$.
	\item The state of a vertex (or an arc) is described by $\mathbf{h}_v$ (or $\mathbf{h}_{vw}$, respectively), i.e., an unknown hidden vector. 
	This concept is analogous to the idea of latent variables present in mixture models or hidden states in hidden Markov models. This state is the subject of learning by the ANN. The trained ANN can then be used to find the hidden state for other graphs.
\end{inparaenum}

When providing the graph representation, our goal is to find an ANN-based model for the variable $y$ describing the whole structure, or for the vector $\mathbf{y}$ of individual variables $y_v$ ($y_{vw}$) related to each vertex (or arc) separately. Such result, known as a readout of the MPNN, represents useful information provided by the ML mechanism for potential prediction, classification etc. To realize this goal, it is necessary to perform the forward pass (inference) with Algorithm~\ref{alg:mpnn} shown below.

\begin{algorithm}
	\caption{Message-Passing Neural Network}\label{alg:mpnn}
	\begin{algorithmic}[1]
		\For{$v \in V$}  
			 \State $\mathbf h_v^0 \leftarrow [\mathbf x_v,0,\ldots, 0]$
		\EndFor
		\For{$t=1$ to $T$} 
			\For{$v \in V$}  
				\State \label{line:m_tilde_v} $\tilde{\mathbf m}_v^{t+1}\leftarrow M_t\left(\mathbf h_v^t,\mathbf h_w^t,\mathbf e_{vw}\right)$ 
			\EndFor
			\For{$v \in V$}  
				\State \label{line:m_v} $\mathbf m_v^{t+1} \leftarrow \sum_{w:(v,w)\in{E}} \tilde{\mathbf m}_v^{t+1} $ \Comment{see line~\ref{line:m_tilde_v}} 
				\State \label{line:h_v} $\mathbf h_v^{t+1} \leftarrow U_t\left(\mathbf h_v^t,\mathbf m_v^{t+1}\right)$ 
			\EndFor
		
		\EndFor
		\State $\hat{\mathbf y} \leftarrow R(\mathbf h)$ 
	\end{algorithmic}
\end{algorithm}

In MPNNs, the inference consists of three operations:
\begin{inparaenum}[\bfseries \itshape (1)]
	\item message-passing,
	\item update, and
	\item finding the readout.
\end{inparaenum}
In iteration $t$, the \emph{message-passing} operation allows the nodes to exchange information described in the form of vector $\mathbf m_v^t$. The \emph{update} operation encodes this information in the hidden state $\mathbf h_v^t$. This process resembles the convergence of a routing protocol, and after a number of iterations (here, indicated by $T$, which is of the order of the average shortest path length), each node holds full information from the whole network. 
These steps represent convergence to the fixed point of a function. 
There are two such functions, $M_t$ and $U_t$. 
At this time point, sparse node and edge features are compressed into a dense representation of the hidden state vectors. This dense representation is fed to the \emph{readout} part of MPNN.

Message-passing and update are exercised alternately $T$ times according to the formulas given in lines~\ref{line:m_v}-\ref{line:h_v} of the Algorithm. 
Both $M_t$ (message function) and $U_t$ (update function) are trainable ANNs. They can be parametrized in different ways, e.g., $M_t$ may also depend on $\mathbf e_{wv}$. Different functions can also be used for incoming and outgoing messages. The hidden representation is initialized as the zero-padded node feature vector: $\mathbf h_v^0=[\mathbf x_v,0,\ldots, 0]$. During the learning process, these zeroes are likely to be replaced with some meaningful values. The readout is obtained from another ANN, and generally equals $\hat{\mathbf y} = R(\mathbf h)$, i.e., it is found on the basis of the stacked node embeddings $\mathbf{h} = [\mathbf h_v]$. 
The simplest readout function preserving basic topological relations between node adjacencies (graph isomorphism) takes the following form:
\begin{equation}\label{eq:r1} \textstyle
	R(\mathbf h) =  f\left(\sum_v \mathbf h_v\right),
\end{equation}
where $f$ is a function approximated by an ANN. Nevertheless, a simple summation is likely to lose a lot of information, and more sophisticated readouts are used in practice~\cite{graph_conv,graph_nn}.



\subsection{MPNN-based Representation of COMNETs}

The approach described in the previous subsection is just one of many possible variations of MPNNs. %
A broad description of different graph neural architectures is presented in~\cite{nnmp:17}. To select a version appropriate to our case, we have to remember that in COMNETs the interest is often focused more on links rather than nodes. Then, MPNN can be defined such that the hidden representation is associated with the link (or link and node) and the \textit{messages} $\mathbf m^t_{vw}$ are sent to connected links (i.e., all the links incident with $v$ and $w$). If their direction is important, the messages can be differentiated between both directions of the arcs. 

As concerns the message-passing operation, we decided to follow~\cite{graph_nn} and use a parametrized affine message function to find $\mathbf m^{t+1}_{v}$:
\begin{equation}\label{eq:afinem} \textstyle
	M_t\left(\mathbf{h}_v^t,\mathbf{h}_w^t,\mathbf{e}_{wv}\right) = A\left(\mathbf{e}_{wv}\right) \times \mathbf{h}_w^t + b\left(\mathbf{e}_{wv}\right),
\end{equation}
where $A$ and $b$ are matrix- and vector-valued ANNs with SELU activations~\cite{Klambauer2017}, respectively. We avoid gradient vanishing by using the \emph{Gated Recurrent Unit}~\cite{Cho2014} for \emph{update} during the learning process: $U_t(\mathbf h_v^t,\mathbf m_v^{t+1}) = \mathit{GRU}(\mathbf h_v^t,\mathbf m_v^{t+1})$. GRU is simpler than the popular LSTM cell, but provides good performance parameters, too~\cite{Chung14a}. Additionally, to reduce the number of parameters we use \emph{weight tying} for all steps $t$, i.e., $U_t=U$, $M_t=M$.

The most complex aspect of our architecture relates to the \textit{readout}. It is obtained with ANN consisting of three parts:
\begin{inparaenum}[\bfseries \itshape (1)]
	\item graph level embedding,
	\item batch normalization layer, and
	\item inference layer.
\end{inparaenum}
The \emph{graph level embedding} $\mathbf R_G$ takes the following form:
\begin{equation}\label{eq:embed} \textstyle
	\mathbf R_G = \sum_v\sigma\left[i\left(\mathbf h_v^T,\mathbf x_v\right)\right] \circ j\left(\mathbf h_v^T\right),
\end{equation}
where $i$ and $j$ are the learned functions (ANNs), and $\circ$ represents the Hadamard product of matrices. 
The ANN $j$ learns to map the node embedding $\mathbf h_v$ into an additive representation, while the ANN $i$ learns to select the most important nodes. The value is mapped with the sigmoid function $\sigma$ into the range $(0,1)$. The result reflects the importance of a given node and the dimension of a hidden representation which is learned during the training process. The calculation of the embedding value is performed such that only the most important node embeddings get multiplied by a number close to $1$ and, in consequence, contribute to the final sum determined by~Eq.~\eqref{eq:embed}. This effect is known as the \emph{attention mechanism}~\cite{ggnn}. 

Having the graph level representation, it is now straightforward to add the final two layers of the ANN. The \emph{batch normalization} layer $z$ improves the training process by learning the average and standard deviation of elements in $\mathbf R_G$ and normalizing them, so that finally we obtain zero means and unit standard deviations~\cite{Ioffe2015}. 
The \emph{inference layer} $f$ provides the final value of the $y$ estimate, so that: $\hat{ y} = f\left[z\left(\mathbf R_G\right)\right]$.

\section{Example: Delays in Queuing Networks}
\label{sec:example}

Here, we show that the proposed approach can be used as a powerful performance evaluation tool providing effective calculation of 
network parameters. The MPNN model can be used to approximate various network performance indicators (e.g., traffic prediction, anomaly detection, delay, $\mathit{RTT}$ estimation). We decided to show the efficiency of the MPNN prediction for average delays in Jackson networks of queues~\cite{Kelly2011}. Contrary to the used example, for real networks the exact relations are frequently unknown or inherently complex, thus in practice the application of ML will considerably simplify network operations. However, in the case of the example presented here, there are a few reasons for using well-known analytical relationships:
\begin{inparaenum}[\bfseries \itshape (a)]
	\item the example is simple and easy to generalize for different domains;
	\item the theoretical formulas for the delay are known, i.e., we can analytically find the values used for training and testing (evaluation) and present a clear interpretation of some aspects of the MPNN used.
\end{inparaenum}


\subsection{Message-Passing Structure for Queuing Networks}

A Jackson network is a network of $M/M/1$ queues with an arbitrary topology. Traffic can enter and leave the network at each node. The external input intensity at node $v$ is given by $\varLambda_v$ and the service rate at that node is denoted as $\mu_v$. Within the network, the traffic is routed according to the routing matrix $[r_{vw}]$. The routing is random and independent, i.e., every packet is routed randomly to the neighboring nodes or leaves the network according to $r_{vw}$. The destination of each packet is independent of the previously taken route. To find the average delay in such a network, we start with the traffic balance equations:
\begin{equation}\label{eq:bilans} \textstyle
	\lambda_v = \varLambda_v + \sum_w r_{wv}\lambda_w \quad v \in V
\end{equation}
The solution to the system given by Eq.~\eqref{eq:bilans} determines the intensity $\lambda_v$ at each node. Then, the average delay $W$ in the network can be computed using Little's law according to the following classical formulas: 
\begin{equation}\label{eq:little} \textstyle
	W = \frac{\sum_v L_v}{\sum_v \varLambda_v} \quad \quad L_v=\frac{\lambda_v}{\mu_v-\lambda_v}, 
\end{equation}
where $L_v$ is  the average queue length in node $v$.

Let us now assume we do not know the relation given by Eq.~\eqref{eq:little}, and want to use the proposed MPNN to learn the delay from the experimental data. To perform such a task, we assume the following regression problem:
\begin{inparaenum}[\bfseries \itshape (a)]
	\item Node features: their traffic intensity and service rates $\mathbf x_v=[\varLambda_v,\mu_v]$.
	\item Edge features are related to routing only: $\mathbf e_{vw}=[r_{vw}]$. 
	\item The predicted readout $y$: the sought network delay $W$. 
\end{inparaenum}

Since the queuing network can be solved analytically, we can provide some insights into why MPNN fits this task. First, let us assume that $\mathbf h_v=[\lambda_v,\varLambda_v,\ldots]$, i.e., the hidden vector explicitly contains  intensities. 
Second, setting $m_v = \sum_w r_{wv}\lambda_w$ allows us to transform Eq.~\eqref{eq:bilans} into a form analogous to 
that given in lines~\ref{line:m_v}-\ref{line:h_v} of Algorithm 1: $m_v^{t+1} = \sum_w  M(\mathbf h_w^t,\mathbf e_{wv})$ and $\mathbf h_v^{t+1}= U(\mathbf h_v^{t},m_v^{t+1})$, with  $M$ and $U$ being functions. In the derivation, we use a fixed point approximation to the solution of~Eq.~\eqref{eq:bilans}, while the nonlinear relationships given by~Eq.~\eqref{eq:little} are approximated by a readout function. 
This way also explains why --- contrary to~\cite{nnmp:17} --- in Eq.~\eqref{eq:afinem} we parametrize $M$ by $\mathbf e_{wv}$, instead of $\mathbf e_{vw}$.

\subsection{Random Networks}


To obtain a sufficient level of confidence of the training results, the MPNN was trained on a high number of different network topologies. We used three types of random networks.
\begin{inparaenum}[\bfseries \itshape (1)]
	\item The first is the Erd{\H{o}}s-R\'{e}nyi (ER) random graph model~\cite{erdos59a}, where two nodes are connected randomly with probability $p$. Since such a construction can produce a disconnected graph, we use the largest connected component of that graph. We set the probability $p=\frac{2}{n}$, where $n = |V|$ is the number of nodes. The training set was created using $n=40$. 
	\item The second model is based on Barab\'{a}si-Albert (BA) random graphs~\cite{Barabasi1999}, where the construction begins with some connected nodes. Each new node is randomly connected to $m$ nodes with connection probabilities proportional to their degrees. This model is known to provide graphs with a long tail nodal degree distribution. We set the number of nodes as a random integer $n \in\interval{10}{40}$, and each new node is initially connected to two other nodes ($m=2$) to represent primary and backup connections.
	\item The last type of topology we tested was based on non-synthetic real networks containing up to 38 nodes. They are taken from the SNDlib collection~\cite{Orlowski10a}, plus the most popular ones: \texttt{janos-us}, \texttt{janos-us-ca}, \texttt{cost266}, \texttt{germany50}. These topologies were retrieved as provided, and used only to evaluate our model.
\end{inparaenum}

As well as topologies, other network parameters were also randomized. The external traffic demands $\varLambda_v$ were randomly sampled from a uniform distribution and normalized to add up to~1. This normalization stabilizes training without loss of generality, since any traffic distribution can be normalized to fall in the range $\interval{0}{1}$ by changing the time unit. The routing matrix $[r_{vw}]$ was generated by assuming an equal probability for each route as well as for leaving the network. In this simple model, nodes of a high degree are more likely to route packets towards the inside of the network. On the other hand, low-degree nodes typically route traffic towards the outside of the network. Using external demands and the routing matrix, intensities at each node $\lambda_v$ were computed using Eq.~\eqref{eq:bilans}. Then, node utilizations were selected randomly from the uniform distribution $\frac{\lambda_v}{\mu_v} \sim \mathcal U(0.3,0.9)$, and the service rate $\mu_v$ was obtained. While the service rate is an independent feature, here we decided to base it on randomized utilizations, since this way we avoid instabilities appearing when $\frac{\lambda_v}{\mu_v}>1$.

The example MPNNs were trained on a collection of $20{,}000$ random graphs. During the training, we used an additional test set of $200$ for periodic testing to avoid overfitting. The final model was evaluated on $2000$ random networks. All three sets were disjoint and topologies were randomly selected to emphasize the structure-independent learning ability of the proposed MPNN model.
The source code is available at~\cite{net2vec}.

\subsection{Discussion of Results}

Evaluation of MPNN models trained on random networks shows that to some extent the model is independent of network topology. The best results were obtained for the BA random graphs in both training and evaluation sets. %
Having said that, it should be stressed that the MPNN trained on the BA model does not generalize as well as the one trained with the ER model. The evaluation results presented in Tab.~\ref{tab:eval} and extended in~\cite{net2vec}  
show that the change from the BA to ER model in the evaluation set results in a substantial systematic error (measured as a mean squared error, $\mathit{MSE}$). However, the Pearson correlation coefficient $\rho$ between predictions and true labels indicates a satisfactory level of dependence.

\begin{table}
	\begin{center}
		\caption{Summary of the obtained evaluation results}
		\label{tab:eval}
		\begin{tabular}{C{1.5cm}|C{1.9cm}|@{}C{1.2cm}@{}|@{}C{1.1 cm}@{}|@{}C{1.15cm}@{}}
			\toprule
			\bfseries Training set model & \bfseries Evaluation set model & $\mathit{MSE}$ & $\mathcal{R}^2$  & $\rho$ \\
			\midrule
			ER        & ER       & 0.0204 & 0.9802 & 0.9937 \\
			ER        & BA       & 0.1251 & 0.9328 & 0.9745 \\
			BA        & BA       & 0.0075 & 0.9923  & 0.9972 \\
			BA        & ER       & 11.62 & --- & 0.8494 \\
			ER        & SNDlib      &  0.0777 & 0.9066 & 0.9752 \\
			ER        & \texttt{janos-us}      &  0.0222 & 0.9434 & 0.9884 \\
			ER        & \texttt{janos-us-ca}      &  0.0454 & 0.9215 & 0.9833 \\
			ER        & \texttt{cost266}      &  0.0374 & 0.9321 & 0.9861 \\
			ER        & \texttt{germany50}      &  0.2131 & 0.7161 & 0.9434 \\
			ER        & ER ($n=60$)      &  0.1420 & 0.9067 & 0.9636 \\
			\bottomrule
			\multicolumn{5}{l}{\scriptsize The table contains the worst limit of 95\% confidence intervals, i.e., a lower} \\
			\multicolumn{5}{l}{\scriptsize bound on $\mathcal{R}^2$ and $\rho$, as well as an upper limit on $\mathit{MSE}$.} \\
			\multicolumn{5}{l}{\scriptsize Confidence intervals: up to 975 permille for $\mathit{MSE}$ and 25 for $\mathcal{R}^2$ and $\rho$}
		\end{tabular}		
	\end{center}
\end{table}

On the other hand, the model trained on the ER networks performs well on both synthetic networks and --- more importantly --- on real topologies. The reason seems to be related to the fact that the ER random graphs cover a much wider range of graph distributions than the BA models. This is an important result, since the former is not perceived as reflecting the structure of existing COMNETs. 

\begin{figure}
	\centering
	\includegraphics[width=\linewidth]{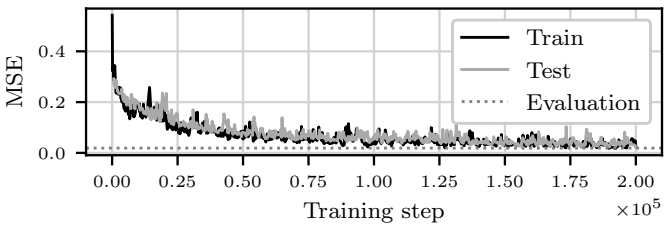}
	\caption{Exponentially smoothed $\mathit{MSE}$ for the ER network model}
	\label{fig:train_er}
\end{figure}

Training characteristics of the ER model are depicted in Fig.~\ref{fig:train_er}. The loss for training and testing is almost the same without any \emph{regularization techniques} being applied. A similar pattern was observed for the BA model, although the final value of $\mathit{MSE}$ was lower. The visible robustness against overfitting is present most likely due to a very simple network architecture consisting of 16 nodes in the network embedding layer. 
Such a small network is capable of generalizing even on larger graphs, i.e., those of sizes never seen during the training process. The values of $\mathit{MSE}$ for the ER networks generated for $n=60$ nodes or \texttt{germany50} SNDlib network ($n=50$) are not as good as for the synthetic evaluation set or SNDlib networks containing $20-38$ nodes (range of most training samples); however, the model provides high percentage values of explained variance $\mathcal{R}^2$ and correlation $\rho$.

It is especially notable that these results are promising from the perspective of applying MPNN for \emph{transfer learning}. In this case, a network once trained on a large dataset of synthetic random topologies can be used as a network feature extractor. The network embedding layer or the final fully connected layer of MPNN can be used as a dense representation of the COMNET relevant to one's application. Hence, this representation can be used as an input for any ML model designed for a particular COMNET. This approach substantially simplifies the training of new models or fine-tuning of models for specific problem domains.

\section{Conclusions and Future Work}
\label{sec:conclusions}

The large variety of COMNET sizes and topologies makes it nontrivial to construct a general machine learning model of sparse graph structured data. In this paper, we show how to apply state-of-the-art graph ANNs (namely, message-passing neural architecture) to simplify learning from existing COMNETs. We train the model on random graphs and obtain the ANN structure, which --- to some extent --- is invariant under COMNET size or topology changes.

Numerical evaluations based on predictions of queuing delays in real COMNET topologies gave the best results when learning was performed with ER random graphs, making them good candidates for training samples in future applications of our approach. %
These applications include network-wise traffic prediction, anomaly detection, and reinforcement learning for network control. The proposed method can be applied to a variety of management cases relevant to contemporary networks, especially those involving programmability (such as Software Defined Networks, SDN) where the power of machine intelligence can be fully embraced.

\section*{Acknowledgments}
  
This work was supported by AGH University of Science and Technology grant, under contract no.~15.11.230.400. The research was also supported in part by PL-Grid Infrastructure.

\ifCLASSOPTIONcaptionsoff
  \newpage
\fi



\end{document}